\documentclass[twocolumn,showpacs,amsmath,amssymb]{revtex4}
\usepackage{amssymb}
\usepackage[dvips]{graphicx}
\begin{document}

\title{Phononic pairing glue in cuprates and related high-temperature superconductors}
\author{A. S. Alexandrov}

\affiliation{ Instituto de Fisica "Gleb Wataghin"/DFA, Universidade Estadual de Campinas-UNICAMP 13083-859, Brasil,  and Department of Physics, Loughborough University,
Loughborough LE11 3TU, United Kingdom\\}

\begin{abstract}
Along with some other researches we have realised that the true origin of high-temperature superconductivity should be found in the strong Coulomb repulsion combined with a significant electron-phonon interaction.  Both interactions are  strong (on the order of 1 eV) compared with the low Fermi energy of doped carries which makes  the conventional BCS-Eliashberg theory  inapplicable in cuprates and related  doped  insulators. Based on our recent analytical and numerical results I argue  that  high-temperature superconductivity from repulsion  is impossible for any strength of the Coulomb interaction. Major steps of our alternative polaronic theory are outlined starting from  the generic Hamiltonian with the unscreened (bare)
Coulomb and electron-phonon interactions accounting for
  critical temperatures of high-temperature superconductors
without any adjustable parameters.

\end{abstract}


\maketitle

\section{High-temperature superconductivity from repulsion: impossibility}
With a few exceptions \cite{special} many researchers share the view that the electron-phonon interaction (EPI) is insufficient for high temperature superconductivity at least in the framework of the conventional Bardeen-Cooper-Schrieffer (BCS) theory or its intermediate-coupling Eliashberg extension \cite{eli}. Phenomenologically, the pairing mechanism of carriers
 could be not only phononic as in the BCS theory  or its
strong-coupling bipolaronic extension \cite{alebook}, but also
excitonic, plasmonic, magnetic, kinetic, or due to some purely repulsive
 interactions combined with the unconventional pairing symmetry of the order
 parameter.

Quite naturally, a number of
authors  assumed that the electron-electron interaction in
high-temperature superconductors was  strong but repulsive providing
high $T_{c}$ without  phonons via  superexchange and/or
spin-fluctuations in unconventional (e.g. d-wave) pairing channel. A
motivation for this concept can be found in the earlier work by Kohn
and Luttinger (KL) \cite{kohn}, who showed that the Cooper pairing
of fermions with any weak  repulsion was possible since the
two-particle interaction induced by many-body effects is attractive
for pairs with  large orbital momenta, $l\gg 1$. While the KL work did not provide the specification of the  actual
angular momentum of condensed Cooper pairs,  Fay and Layzer
\cite{fay} found that a system of \emph{hard-sphere} fermions
condenses at low densities into p-orbital state ($l=1$). The
critical transition temperature  T$_{c}$ of repulsive fermions was
estimated well below 0.1 K with very little enhancement due to
flatness of the Fermi surface \cite{lut}.  In two dimensions (2D)
the KL effect is absent for the parabolic band dispersion, but the
d-wave low-T$_c$ pairing was found with the repulsive Hubbard $U$ (i.e. hard-sphere) potential when
tight-binding corrections to the electron energy spectrum are taken
into account \cite{bar}.

More recent studies  claimed that even a weak repulsive Hubbard $U$
 combined with lattice induced band-structure effects
results in higher  values of T$_c$
 in the spin singlet d-wave channel near half-filling "encouragingly
similar to what is found in the cuprate high-temperature
superconductors" \cite{kivsca,kiv}.

\begin{figure}[tbp]
\begin{center}
\includegraphics[angle=-90,width=0.55\textwidth]{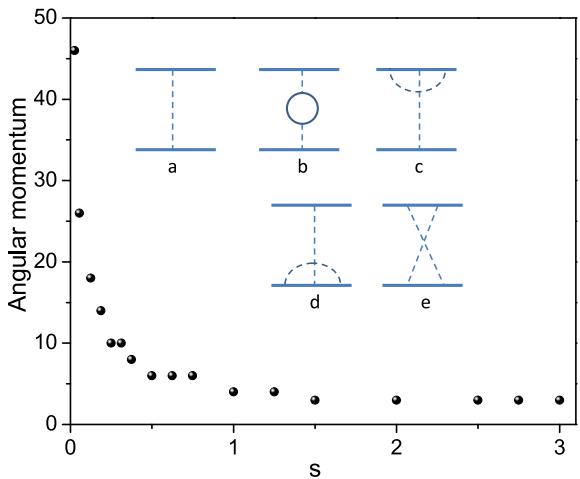}
\end{center}
\caption{(Color online) Angular momentum of the order
parameter as the function of the dimensionless repulsion $s=e^2/\pi \hbar v_F$ in the 3D Coulomb gas. Inset shows
the first order (a) and the second-order (b,c,d,e) contributions to the
two-particle vertex. (Reproduced from
Ref. \cite{alekab}, \copyright  American Physical Society,
2011.)} \label{one}
\end{figure}

Different from
these studies we have analysed the KL problem with the realistic Coulomb repulsion rather than with the hard-core Hubbard $U$ \cite{alekab}.
The second-order diagrams, responsible for the unconventional pairing in the KL theory Fig.(\ref{one}),  are prohibitively
difficult to evaluate when the Fourier transform of the repulsive
potential, $v(\textbf{q})$ depends on $q$, so that most previous and
recent studies \cite{kivsca,kiv} confined to the hard-sphere
Hubbard $U$ repulsion with $v(\textbf{q})$=constant. In that case
the first order does not contribute to the pairing vertex in any
unconventional channel with $l\geqslant 1$, and the only
(attractive) contribution comes from the second-order exchange diagram providing p-wave pairing in 3D for parabolic
dispersion \cite{fay} and d-wave pairing in 2D for the
 tight-binding dispersion \cite{bar,kivsca,kiv}.

Actually the screening length in the  dense Coulomb
gas, where the perturbation expansion in powers of $s=e^2/\pi \hbar v_F$ makes
sense,  is \emph{large} compared with the characteristic electron wavelength $\sim 1/k_F$, so that
it is  unreasonable to treat  the weak Coulomb repulsion
 as short-ranged as in Ref.\cite{kiv} ( $v_F$ and $k_F$ are the Fermi speed and the wave-vector, respectively). We have
 included all buble diagrams in the screened Coulomb potential, $v(q)=4\pi e^2/[q^2+s(q)]$ with the static Lindhard function,
\begin{equation}
s(q)=4sk_F^2\left[{1\over{2}}+{k_F^2-q^2/4
\over{2q k_F}}\ln {k_F+q/2\over{k_F-q/2}}\right],
\end{equation}
 and carefully calculated all second-order diagrams in the 3D Cooper channel \cite{alekab}.
Then  solving the linearised BCS equation  allowed us to find the symmetry (i.e. the angular momentum $l$) of the
order parameter, shown
in Fig.(\ref{one}) as the function of the interaction strength $s$. Remarkably we did not observe p
and d-wave pairing in the whole region  of the Kohn-Luttinger
perturbation expansion and beyond (up to $s=3$). There is a pairing in higher momentum
states, $l \geqslant 3$, but the corresponding eigenvalues  are
numerically so small ($\lambda_3\approx 0.0011$ at $s=3$), that
the corresponding T$_c \approx E_F \exp(-1/\lambda)$ is virtually zero for any realistic Fermi
energy $E_F$.

To analyse the unconventional pairing in  a two-dimensional (2D) electron gas on
the square lattice  with the tight-binding energy dispersion,
\begin{equation}
E_\textbf{p}=-2t [\cos(p_x a/\hbar)+\cos(p_y a/\hbar)]-\mu
\end{equation}
($a$ is the lattice constant) we  used the model 2D Coulomb potential $v(q)=2\pi e^2/(q+\kappa)$ with a constant inverse screening length $\kappa$.   For the half-filled band the Fermi
level, $\mu$ is found at the van-Hove singularity (vHs) of the
density of states, $\mu=0$, so that one might expect a strong
enhancement of the unconventional T$_c$ near half-filling due to
vHs proximity \cite{kivsca,kiv}. Numerically solving the linearised BCS equation by
discretization of the Fermi surface  we reproduced fairly well the
results of Ref.\cite{kivsca} for the 2D Hubbard model, where the ground state is $B_{1g}$ spin singlet with the d-wave symmetry $x^2-y^2$ close to the half-filling. However,
 using the screened Coulomb repulsion instead of the Hubbard one qualitatively  changed the ground state, so that
 contrary to Ref. \cite{bar,kivsca,kiv} neither p- nor d-wave  pairing is possible  at any filling while the repulsion is weak \cite{alekab}. This surprising result is due to a
nonvanishing first-order repulsive  contribution to  unconventional channels from the finite range interaction.

In cuprate and related superconductors  the
Coulomb repulsion is believed  rather strong $s \gg 1$, so that the
perturbative KL approach might have no direct relevance to these
materials. Different numerical techniques have been applied to
elucidate the ground state of the repulsive Hubbard model in the
intermediate to strong-coupling regime, sometimes with
conflicting conclusions. In particular, recent studies by Aimi and
Imada \cite{imada}  using a sign-problem-free Gaussian-basis Monte
Carlo (GBMC) algorithm  showed that the simplest Hubbard model
with the nearest-neighbor hopping has no superconducting
condensation energy   at optimum doping. This striking result has
been confirmed in the variational Monte Carlo (VMC) studies by
Baeriswyl \emph{et al.} \cite{bae}, who found, however, some
condensation energy away from  optimum doping and also adding
next-nearest neighbor hopping.

\begin{figure}
\includegraphics[width = 63mm, angle=-90]{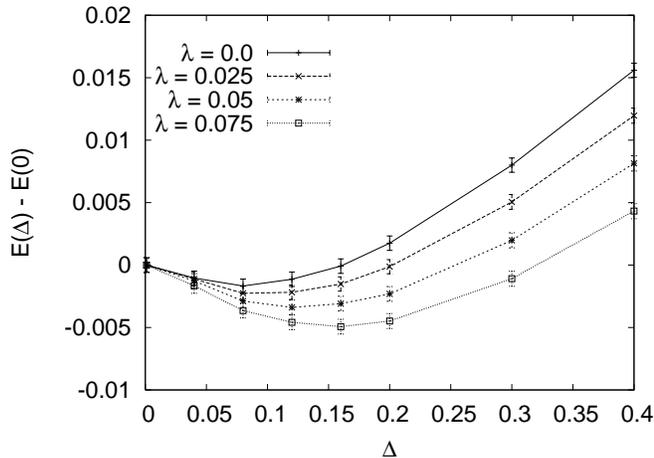}
\caption{Condensation energy per electron (in units of the hopping integral $t$)
versus the amplitude of the superconducting d-wave order-parameter
in the 2D Hubbard model with the finite-range EPI for $U/t=8$,
and different EP coupling $\lambda$. (Reproduced from
Ref. \cite{hardy}, \copyright  American Physical Society,
2009.)} \label{DWave}
\end{figure}

Using   similar VMC simulations of the Hubbard model but including a reaslistic finite-range electron-phonon interaction,
we  found that even a relatively weak EPI
 with the BCS coupling constant $\lambda \approx 0.1$  induces the d-wave
superconducting state in the model with the
condensation energy several times larger than can be obtained with the
Hubbard repulsion alone \cite{hardy}, Fig.(\ref{DWave}). Moreover, the unconventional
superconductivity has been shown to exist due to a finite-range
EPI \cite{hardy,alesym} without the need
for additional mechanisms such as spin fluctuations.

Based on these findings one concludes  that the p- and d-wave Cooper
pairing from the weak  Coulomb repulsion  is not possible  between
fermions  in any dimension.  Pairing in
higher momentum states  $(l \geqslant 3)$  has  the corresponding
T$_c$ virtually zero for any realistic Fermi energy. The
unconventional pairing from the strong Coulomb repulsion is highly unrealistic  either since the corresponding condensation energy, if
any, is many times lower than the condensation energy caused by the realistic
 electron-phonon interaction.

\section{Alternative theory}
Based on the above analytical and numerical results  and numerous experimental  facts supporting strong electron-lattice coupling \cite{alerev} we trust that EPI should be  the key in solving the high-T$_c$ problem. However, a quantitative
analysis of the doping-dependent EPI has remained elusive in cuprates and related compounds.
 Recent observations of the quantum magnetic
oscillations in some underdoped \cite{undo} and overdoped \cite{overdo}
cuprate superconductors has opened up the possibility for
a quantitative assessment of EPI in these and related doped
ionic lattices \cite{alebra}. The oscillations revealed cylindrical Fermi
surfaces, enhanced effective masses of carriers (ranging
from $2m_e$ to $6m_e$) and the astonishingly low Fermi
energy, which appears to be well below 40 meV
in underdoped  \cite{undo} and less than or about
400 meV in heavily overdoped cuprates \cite{overdo}. Such low Fermi energies
 make the Migdal-Eliashberg  adiabatic approach
to EPI  inapplicable in these compounds. Since
carriers in cuprates are in the non-adiabatic (underdoped) or
near-adiabatic (overdoped) regimes  with their Fermi energy about the optical phonon energy, our multi polaron theory of superconductivity \cite{alebook} should be more appropriate here.

The theory of dense polaronic systems in the intermediate
coupling regime has remained rather cumbersome, in particular, when
EPI competes with strong electron correlations. Corresponding
microscopic models with the on-site Hubbard repulsion and the
short-range Holstein EPI have been studied using analytical and powerful numerical
techniques \cite{alerev}.

In most analytical and numerical studies
 both interactions have been introduced as
input parameters not directly related to the material.  Quantitative
calculations of the interaction matrix elements can be performed
from pseudopotentials using the density functional theory (DFT) \cite{bauer}. On
the other hand, one can express the bare Coulomb repulsion and EPI
through material parameters rather than computing them from first
principles in many physically important cases \cite{mahan}. In
particular, for a polar coupling to longitudinal optical phonons (the Fr\"{o}%
hlich EPI), which is the major EPI in polar crystals, both the momentum
dependence of the matrix element, $M(\mathbf{q})$, and its magnitude are
well known, $|M(\mathbf{q})|=\gamma (q))\hbar \omega _{0}/\sqrt{2N}$. Here  $\gamma (q)=\sqrt{4\pi e^{2}/\kappa \Omega \hbar \omega
_{0}q^{2}}$ is a
dimensionless coupling,  $\Omega $ is a unit cell volume, $N$ is the number of
unit cells in a crystal, $\omega _{0}$ is the optical phonon frequency, and $%
\kappa =\epsilon _{\infty }\epsilon _{0}/(\epsilon _{0}-\epsilon _{\infty })$
($\epsilon _{\infty }$ and
$\epsilon _{0}$ are high-frequency and static dielectric constants  in a parent
polar insulator).

Recently I have noticed that, in  highly polarizable ionic
lattices as cuprates with $\epsilon _{0}\gg 1$ the \emph{bare long-range} Coulomb and  Fr\"{o}%
hlich
interactions almost negate  each other allowing for the analytical multi-polaron theory  in the strong-coupling regime \cite{Alexandrov2011}. The
dielectric response function of strongly correlated electrons is \emph{%
apriori} unknown. Hence one has to start here with a generic
Hamiltonian including \emph{unscreened} Coulomb and Fr\"ohlich
interactions operating on the same scale since any ad-hoc assumption
on their range and relative magnitude might fail,
\begin{eqnarray}
H &=-&\sum_{i,j} (T_{ij}\delta_{ss^\prime}+\mu \delta _{ij}) c_{i}^{\dagger
}c_{j} +{\frac{1}{{2}}}\sum_{i\neq j}{\frac{e^2}{{\epsilon_\infty |\mathbf{%
m-n}|}}}\hat{n}_{i}\hat{n}_j+\cr && \sum_{\mathbf{q},i}\hbar \omega _{0}\hat{%
n}_{i}\left[ u(\mathbf{m,q} ) d_{\mathbf{q} }+H.c.\right]+H_{ph}.
\label{hamiltonian}
\end{eqnarray}
Here $T_{ij}\equiv T(\mathbf{m-n})$ is the bare hopping integral, $\mu$ is
the chemical potential, $i=\mathbf{m},s$ and $j=\mathbf{n},s^{\prime }$
include both site $(\mathbf{m,n})$ and spin $(s,s^{\prime })$ states, $u(%
\mathbf{m,q})= (2N)^{-1/2}\gamma(q)\exp(i \mathbf{q \cdot m})$, $c_{i}, d_{%
\mathbf{q} }$ are electron and phonon operators, respectively, $\hat{n}%
_{i}=c^\dagger_i c_i$ is a site occupation operator, and $H_{ph}=\sum_{%
\mathbf{q}}\hbar \omega _{0}(d_{\mathbf{q} }^{\dagger }d_{\mathbf{q}}+1/2)$
is the polar vibration energy.

This Hamiltonian
can be reduced to the polaronic "t-J$_p$" Hamiltonian using
 two successive canonical transformations \cite{Alexandrov2011},
\begin{eqnarray}
\mathcal{H}&=&-\sum_{i,j}(t_{ij}\delta_{ss^\prime}+\tilde{\tilde{\mu}}
\delta_{ij})c_{i}^{\dagger }c_{j} \cr &+&2 \sum_{\mathbf{m} \neq \mathbf{n}}
J_p(\mathbf{m}-\mathbf{n}) \left(\vec{S}_\mathbf{m} \cdot \vec{S}_\mathbf{n}+%
{\frac{1}{{4}}}\hat{n}_\mathbf{m}\hat{n}_\mathbf{n}\right),  \label{tJ}
\end{eqnarray}
where $\vec{S}_\mathbf{m}=(1/2)\sum_{s,s^\prime}c^\dagger_{\mathbf{m}s}\vec{%
\tau}_{ss^\prime} c_{\mathbf{m}s^\prime}$ is the spin 1/2 operator ($\vec{%
\tau}$ are the Pauli matrices), $\hat{n}_\mathbf{m}=\sum_s \hat{n}_i$, and $%
\tilde{\tilde{\mu}}$ is the renormalised chemical potential.

There is a striking difference between this polaronic t-J$_p$
Hamiltonian Eq.(\ref{tJ}) and the familiar t-J model derived from
the repulsive Hubbard U Hamiltonian in the limit $U\gg t$ omitting
the so-called three-site hoppings and EPI \cite{tJ}. The latter model acts in
a projected Hilbert space constrained to no double occupancy. Within
this standard t-J model the bare transfer amplitude of electrons
($t$) sets the energy scale for incoherent transport, while the
Heisenberg interaction ($J \propto t^2/U$) allows for spin flips
leading to coherent hole motion with an effective bandwidth
determined by $J \ll t$.  On the contrary in our polaronic t-J$_p$
Hamiltonian, Eq.(\ref{tJ}) there is no constraint on the double
on-site occupancy since the Coulomb repulsion is negated by the
Fr\"ohlich EPI. The polaronic hopping integral $t$ leads to the
coherent (bi)polaron band and the antiferromagnetic exchange of
purely phononic origin $J_p \gg t$  bounds polarons into small
superlight inter-site bipolarons. Last but not least the difference
is in the "+" sign in
the last term of Eq.(\ref{tJ}) proportional to $\hat{n}_\mathbf{m}\hat{n}_%
\mathbf{n}$, which protects the ground superconducting state from
the bipolaron clustering, in contrast with the "-" sign in the
similar term of the standard t-J model, where the phase separation
is expected at sufficiently large $J$.

The reduction  of Eq.(\ref{hamiltonian}) to Eq.(\ref{tJ}) is based on the so-called
"inverse coupling constant ($1/\lambda$)" perturbation technique
developed by us for the multi-polaron systems
\cite{alebook}, where the residual polaron-phonon interaction
creates multi-phonon vertexes in the diagrammatic technique. Different from any model proposed so far, all quantities in the
polaronic $t$-$J_{p}$ Hamiltonian \eqref{tJ} are
defined through the  material parameters, in
particular $t_{ij}=T(%
\mathbf{m}-\mathbf{n})\exp[-g^2(\mathbf{m}-\mathbf{n})]$ with
\begin{equation}\small
g^2(\mathbf{m})={\frac{2 \pi e^2}{{\kappa \hbar \omega_0
N\Omega}}}\sum_\mathbf{q}{1-\cos(\mathbf{q }\cdot \mathbf{m})\over{q^2}},
\label{t}
\end{equation}
and
\begin{equation}\small
J_p(\mathbf{m})={T^2(\mathbf{m})\over{2g^2(\mathbf{m}) \hbar
\omega_0}}, \label{J}
\end{equation}
Here the high-frequency, $\epsilon _{\infty }$ and the static,
$\epsilon _{0}$ dielectric constants as well as  the optical phonon
frequency, $\omega_0$ and the bare hopping integrals  in a rigid
lattice, $T(\mathbf{m})$
 are  measured and/or found using the first-principle Density Functional Theory \cite{bauer} in a parent polar insulator.

 Solving  the
polaronic $t$-$J_{p}$ model with the appropriate material parameters, $\kappa, \omega_0$ and $T(\mathbf{m})$ exhibits a phase transition to a
superconducting state with the critical temperature  in excess of
$100$K at optimum doping \cite{Alexandrov2011,sica}. The model provides un understanding of the spin and charge pseudogaps \cite{sica} and the unified parameter-free explanation of the observed oxygen-isotope efects on the critical temperature, the magnetic-field penetration depth, and on the normal-state pseudogap in
underdoped cuprate superconductors \cite{alepet}.

I highly appreciate fruitful collaboration with Alexander Bratkovsky, Viktor Kabanov, John Samson, Gerardo Sica and Peter Zhao, and stimulating discussions with   Annette
Bussmann-Holder, Andrey  Chubukov,  Jorge Hirsch, Maxim Kagan, Hugo Keller, Steven Kivelson,  Viktor Khodel, Frank
Marsiglio, Dragan Mihailovic and Roman Micnas.


\begin{thebibliography}{90}
\bibitem{special}  E. G. Maksimov, M. L. Kulic', and O. V. Dolgov,
 Advances in Condensed Matter Physics Volume 2010 (\emph{Special Issue on Phonons and Electron Correlations in High-Temperature and Other Novel Superconductors}), Article ID 423725 (2010).
\bibitem{eli}  G. M. Eliashberg, Zh. Eksp. Teor. Fiz. \textbf{39}, 1437 (1960)
[Sov. Phys. JETP  \textbf{12}, 1000 (1960)].
\bibitem{alebook} A. S. Alexandrov, \emph{Theory of Superconductivity: From Weak
to Strong Coupling} (IoP Publishing, Bristol  2003).
\bibitem{kohn} W. Kohn  and J. M.  Luttinger, Phys. Rev. Lett. \textbf{15},
524 (1965).
\bibitem{fay} D. Fay  and A. Layzer, Phys. Rev. Lett. \textbf{20}, 187
(1968).
\bibitem{lut}  J. M. Luttinger, Phys. Rev. \textbf{150}, 202 (1966)
\bibitem{bar} M. A. Baranov, A. V. Chubukov and M. Yu. Kagan, Int.
J. Mod. Phys. \textbf{14}, 2471 (1992).
\bibitem{kivsca} S.  Raghu,  S. A. Kivelson, and D. J.  Scalapino, Phys. Rev. B  \textbf{81 },
224505 (2010).
\bibitem{kiv} S. Raghu and  S. A. Kivelson, Phys. Rev. B \textbf{83}, 094518
(2011).

\bibitem{alekab} A. S. Alexandrov and V.V. Kabanov, Phys. Rev. Lett. \textbf{106}, 136403 (2011).
\bibitem{imada} T. Aimi and M. Imada, J. Phys. Soc. Jpn. {\bf 76}, 113708
(2007).
\bibitem{bae} D. Baeriswyl, D. Eichenberger and  M. Menteshashvili, New J. Phys. \textbf{11}, 075010 (2009).
\bibitem{hardy}
T. M. Hardy, J. P. Hague, J. H. Samson, and A. S. Alexandrov,
Phys. Rev. B \textbf{79}, 212501 (2009).
\bibitem{alesym} A. S. Alexandrov, Phys. Rev. B \textbf{77}, 094502
(2008).
\bibitem{alerev} for a review see A. S. Alexandrov  and J. T.  Devreese, \emph{Advances in
Polaron Physics }(Springer, Berlin 2009).
\bibitem{undo} N. Doiron-Leyraud et al., Nature (London)\textbf{ 447}, 565
(2007); E.A. Yelland et al., Phys. Rev. Lett. \textbf{100},
047003 (2008); C. Jaudet et al., Phys. Rev. Lett. \textbf{100},
187005 (2008).
\bibitem{overdo}  B. Vignolle et al., Nature (London)\textbf{ 455}, 952 (2008); A.F.
Bangura et al., Phys. Rev. B \textbf{82}, 140501(R) (2010).

\bibitem{alebra}  A. S. Alexandrov  and A. M.  Bratkovsky, Phys. Rev. Lett.
\textbf{105}, 226408 (2010).
\bibitem{bauer} T. Bauer and C. Falter, Phys. Rev. B \textbf{80}, 094525 (2009).
\bibitem{mahan} G. D. Mahan, \emph{Many-Particle Physics} (Plenum, New York
1990).
\bibitem{Alexandrov2011} A. S. Alexandrov, EPL  \textbf{95},  27004 (2011).
\bibitem{tJ} J. E. Hirsch, Phys. Rev. Lett. \textbf{54}, 1317 (1985); J. Spalek,
Phys. Rev B \textbf{37}, 533 (1988);
\bibitem{sica} A. S. Alexandrov, J. H. Samson, and G. Sica, arXiv:1112.3230.
\bibitem{alepet} A. S. Alexandrov and G. M. Zhao, to appear in New J. Phys. \textbf{13} (2011).



\end{thebibliography}
\end{document}